\documentclass[journal]{IEEEtran}
\usepackage[T1]{fontenc}
\usepackage{amsmath,amssymb,booktabs,graphicx,url,cite}
\newtheorem{proposition}{Proposition}
\begin{document}

\title{Lossless Re-Encoding of Vorbis and Opus Audio Files}

\author{Kamila~Szewczyk%
\thanks{The author is with the Algorithmic Bioinformatics group, Department of Computer Science, and with the Department of Mathematics, Saarland University, Saarbr\"ucken, Germany (e-mail: szewczyk@cs.uni-saarland.de).}}

\maketitle

\begin{abstract}
We present Balrogg, a lossless recompressor that replaces the entropy coding of Ogg Vorbis and Ogg Opus files with adaptive context mixing models. For Vorbis, we write each represent each codebook entry as a tuple of small signed integers (digits), binarized in five stages, and mixed with a direct context model, a bit-history model, and a match model, with weights selected by the neighboring digits and the previous residue pass. For Opus, each symbol is coded by an adaptive context model initialized with the codec's probability tables as priors, and the pyramid vector quantizer (PVQ) indices of the CELT layer are split recursively into pulse counts. At the default level, Balrogg reduces Vorbis music by 16.2\%, Vorbis speech by 10.6\%, and Opus speech by 5.4\%; xz, zstd, ZPAQ, and bzip3 save at most 4.1\% on the same files. Over 48 encoder configurations, files shrink to 66.5--96.1\% of their size. Encoding and decoding are symmetric, run at 1.8--3.2~MB/s per core, and use less than 12~MiB of memory.
\end{abstract}


\section{Introduction}
\IEEEPARstart{V}{orbis} \cite{vorbis} and Opus \cite{opus} use simple, fast entropy coders. Vorbis codes vector quantizer codebook entries with static Huffman codes. Opus uses a range coder with fixed probability tables and codes the pyramid vector quantizer (PVQ) indices of its CELT layer \cite{fischer,opusmusic} with a uniform distribution. Neither codec adapts its entropy code to the statistics of a recording.

A lossless recompressor undoes the entropy coding and codes the symbols again with stronger models. The decoder restores the original file, so checksums and signatures remain valid; this suits archival and transmission.

packMP3 \cite{packmp3} is prior art on Mp3 recompression. packJPG and Lepton \cite{packjpg,lepton} losslessly reduce JPEG files by about 20--23\%. OptiVorbis \cite{optivorbis} recomputes the Huffman codes of a Vorbis file and writes a smaller playable file, which differs from the input (it is thus, in some sense, the dual of lossless recompression). On the bit-packed Vorbis and Opus streams of Section~\ref{sec:results} \textit{prima facie}, general-purpose compressors save only about 0.1--4.1\% of the file size.

This paper describes Balrogg\footnote{Balrogg is free software under the GNU GPLv3: \url{https://github.com/iczelia/balrogg}. All statements and measurements refer to tag \texttt{1.5} (commit \texttt{c7aedca}).}, to our knowledge the first lossless recompressor for Vorbis and Opus. Section~\ref{sec:structure} outlines the coder. Section~\ref{sec:vorbis} describes the Vorbis coding model: a codebook-independent representation of residue entries and an online logistic mixer whose weights are selected by the neighboring digits and the previous residue pass. Section~\ref{sec:opus} describes the Opus models: adaptive models with pre-seeded prior and a bound on their regret (Proposition~\ref{prop:regret}); a recursive split of PVQ indices that keeps the codec's code length for any split tree (Proposition~\ref{prop:tree}). Section~\ref{sec:results} reports size, speed, and memory on five data sets, 48 encoder configurations, and random samples from the Internet Archive. We assume familiarity with binary context mixing \cite{mahoney,knoll,mattern}, range coding \cite{martin}, and the Vorbis and Opus bitstreams.

\section{Structure of the Coder}
\label{sec:structure}
The encoder splits the Ogg pages \cite{ogg,oggopus} into packets and framing and parses each packet up to its entropy-coding stage. No transform is computed. This gives three kinds of data:
\begin{enumerate}
\item payload symbols, coded by the models of Sections~\ref{sec:vorbis} and~\ref{sec:opus};
\item container data (page headers, lacing values, and the identification and comment headers), coded by residual and byte models;
\item side information: Vorbis classwords that differ from their prediction, nonzero padding, truncated packets, and Opus frame bytes that differ from the output of the reference range encoder (Section~\ref{sec:frame}).
\end{enumerate}

Binary decisions use a 32-bit range coder with carry propagation (after D.~Subbotin) and 16-bit probabilities. Opus multisymbol decisions use a 32-bit range coder with integer cumulative frequencies totalling at most $2^{17}$. The archive multiplexes numbered streams in chunks of at most 1~MiB without resetting the coders. All models use integer arithmetic only. The archive stores no model parameters except the three tuning bytes of Section~\ref{sec:effort} and does not depend on the compiler, word size, byte order, or vector instruction set.

\section{Vorbis}
\label{sec:vorbis}
\subsection{Residue digits}
A Vorbis codebook of lookup type~1 and dimension $d$ has integer multiplicands $M[0],\ldots,M[n_v-1]$. It maps the entry
\begin{equation}\label{eq:digits}
 e=\sum_{j=0}^{d-1}a_jn_v^j,\qquad 0\le a_j<n_v,
\end{equation}
to the vector $(M[a_0],\ldots,M[a_{d-1}])$, which the decoder scales and adds to the residue. Balrogg codes the $d$ \emph{digits}
\begin{equation}\label{eq:center}
 v_j=M[a_j]-b,\qquad b=\Bigl\lfloor\frac{1+\max_aM[a]}2\Bigr\rfloor.
\end{equation}
Centering puts the digits of all codebooks, whatever their step size and codeword lengths, on one integer grid, so statistics learned on one codebook transfer to others. The decoder inverts \eqref{eq:center} with $M^{-1}$, computes $e$ by \eqref{eq:digits}, and writes the codeword of $e$.

Similar codebooks share models. Two codebooks are assigned the same model slot if they have the same number of entries, the same number of multiplicands, and the same quantizer step, and if their codeword lengths differ by at most one bit per used entry on average. Slots persist across chained links, so a link that redefines a known codebook reuses its models.

\subsection{Binarization and direct models}
A digit $v$ is coded as binary decisions in up to five stages: (0) whether $v\ne0$; (1) the sign of $v$; (2) whether $|v|=1$; (3) $\ell=\lfloor\log_2|v|\rfloor-1\in\{0,\ldots,7\}$, as a three-bit binary tree; (4) the $\ell+1$ bits of $|v|$ below its leading one. Coding stops after stage~0 if $v=0$ and after stage~2 if $|v|=1$. The stages cover $|v|<2^9$; libvorbis writes $|v|\le54$ and FFmpeg $|v|\le8$. The \emph{magnitude class} of $v$ is the bit length of $|v|$: $0$ for $v=0$, $1$ for $|v|=1$, and $\ell+2$ otherwise.

Each decision has a direct context model; i.e., an adaptive (EMA) estimator in a table indexed by the context of Table~\ref{tab:ctx}. The estimator holds a 16-bit probability $P$ that the bit is zero and a count $n$. Let $r_n=\lfloor65535/(n+3)\rfloor$. After the bit $z$ is coded,
\begin{equation}\label{eq:bitupd}
 P\gets\begin{cases}
 P+\lfloor(65535-P)\,r_n/2^{16}\rfloor,&z=0,\\
 P-\lfloor P\,r_n/2^{16}\rfloor,&z=1,
 \end{cases}
\end{equation}
and $n$ is incremented up to a limit $n_{\max}$ (120 by default). The rate thus decays as $1/(n+3)$ down to $1/(n_{\max}+3)$. Level~1 uses the direct models alone.

\begin{table}[t]
\centering
\caption{Contexts of the direct models of a residue digit. $c$: position, $\lambda(c)$: bit length of $c$, $h$: channel (clipped to 3), $m$: memory, $g$: magnitude class of the latest digit of the first channel, $f=[g\ne0]$, [slot unchanged]: the previously coded digit used the same slot.}
\label{tab:ctx}
\setlength{\tabcolsep}{4pt}
\begin{tabular}{@{}ll@{}}\toprule
Stage&Context\\\midrule
0, $v\ne0$&$\lfloor c/4\rfloor$, $h$, $f$, $[m=0]$, [slot unchanged]\\
1, sign&$\lambda(c)$, $h$, $f$, $[m\ne0]$, sign of $m$, two last signs in $h$\\
2, $|v|=1$&$\lambda(c)$, $h$, $g$, $[m\ne0]$, sign of $v$\\
3, $\ell$&$h$, $g$, node of the three-bit tree\\
4, mantissa&$h$, $g$, $\ell$, two last mantissa bits\\\bottomrule
\end{tabular}
\end{table}

Let $c$ be the position of a digit in its channel $h$, after de-interleaving for residue type~2. Besides $c$ and $h$, the contexts use two earlier digits.

The first is the latest digit of the first channel, with magnitude class $g$ and $f=[g\ne0]$. Since a stereo type-2 residue interleaves the channels, this is the other channel's digit at the same position for $h=1$, and the same channel's digit at the previous position for $h=0$.

The second is the digit at the same position in the previous residue pass. Per residue, pass, channel, and position, the coder keeps one byte holding the sign and magnitude class of the last digit coded there. The \emph{memory} $m$ of a digit in pass $\tau>0$ is the record of pass $\tau-1$ at its position; in the first pass, it is the record left by the first pass of an earlier packet.

\subsection{Mixing}
From level~2 on, an online logistic mixer \cite{mahoney,knoll,mattern} computes the probability of a decision from six inputs $t_1,\ldots,t_6$. Let $\operatorname{st}(p)=\ln\frac p{1-p}$ and $\operatorname{sq}=\operatorname{st}^{-1}$. With a weight vector $w$, selected by context as described below, the probability of a one bit is $p=\operatorname{sq}(\sum_iw_it_i)$. After the bit $y$ is coded,
\begin{equation}\label{eq:mixupd}
 w_i\gets w_i+\eta\,2^{-11}(y-p)\,t_i,
\end{equation}
with the learning rate $\eta=7$ by default. The inputs are:
\begin{enumerate}
\item $t_1=\operatorname{st}(1-P)$, the prediction of the direct model;
\item $t_2=\operatorname{st}(p_s)$, the prediction of a bit-history model. Each stage has a table of $2^{18}$ one-byte states, indexed by a hash of the slot, $h$, $c$, the sign in $m$, and the two low bits of the magnitude class in $m$. An adaptive state map gives the probability $p_s$ of a one bit in state $s$;
\item $t_3=1$, a bias;
\item $t_4=\varepsilon t_2$, with $\varepsilon=+1$ if the history $s$ contains only zeros, $-1$ if it contains only ones, and $0$ otherwise;
\item $t_5=\pm\operatorname{st}(\alpha)$, where $\alpha$ is the probability that the match model predicts the bit correctly, learned for each stage and match length;
\item $t_6=\pm\min(L,32)/4$, where $L$ is the match length.
\end{enumerate}
The states are the 253 states of \cite{mahoney}, which count small numbers $(n_0,n_1)$ of zeros and ones exactly and large numbers approximately. The state map starts at $(n_1+1)/(n_0+n_1+2)$, with a more confident value for histories of a single bit value, and is updated by \eqref{eq:bitupd}.

The mixer differs from those of general-purpose context-mixing compressors in three ways.

\begin{itemize}
\item \emph{Weight selection.} Each stage has 200 weight vectors. A vector is selected by four values: (i) the magnitude class, clipped to~4, of the preceding digit of the same channel, or 0 if that digit is not at position $c-1$; (ii) the clipped magnitude class of the most recently coded digit if it belongs to another channel, and 0 otherwise; (iii) the parity of $h$; (iv) the two low bits of the magnitude class in $m$.

\item \emph{Initialization.} Every weight vector is initialised to $(1,0,\ldots,0)$, so the mixer initially reproduces the direct model, i.e., the level-1 code, and the other inputs gain weight only when they improve on it. This prevents underfitting in the ``curse of dimensionality'' for short files.

\item \emph{Digit-level match model.} The match model predicts whole digits, not bytes or bits. It stores the last $2^{18}$ digits, clipped to $[-127,127]$, and a table of $2^{16}$ positions indexed by a rolling hash of the last ten digits. A candidate is verified by comparing the digits that precede it. If at least ten digits agree, the digit that follows the candidate is the prediction, and the match length $L$ counts up to 63. Each stage receives the bit implied by the predicted digit, up to the first mismatch. The match model is off by default; the search of Section~\ref{sec:effort} may enable it for a file.
\end{itemize}

The mixer uses 16-bit fixed-point arithmetic. Log-odds are stored in units of $1/256$ and clamped to $|x|<8$, weights are stored in units of $2^{-15}$ and saturate, and the error $y-p$ has 12 bits. The function $\operatorname{sq}$ is tabulated at 33 points and interpolated linearly. $\operatorname{st}$ is the tabulated inverse of $\operatorname(sq)$.

\subsection{Floors, classes, and exceptional packets}
The values of a type-1 floor are coded in order of ascending frequency. Each value is coded as a zero flag, a bit length, and mantissa bits, all with the floor number and the post as context. In addition, the zero flag uses the two previous flags, the bit length (a three-bit tree) uses the previous bit length at the same post, and a mantissa bit uses its position and the two previous bits.

Several classwords can encode the same floor partition. The coder predicts the one that selects, for each value, the first subclass codebook able to represent it, and codes a classword only when it differs from the prediction.

The class of a residue partition is coded as a four-bit tree with the partition index and the residue number as context. If the search of Section~\ref{sec:effort} enables it, the context also includes the class of the same partition in the preceding packet, reduced to six buckets.

A pre-pass over each packet detects four exceptional cases: a mispredicted classword, nonzero padding, trailing bits, and truncation. A packet-type symbol flags them. Trailing bits, including a final partial byte, are coded separately, and a truncated packet is coded by an adaptive byte model.

\subsection{Levels and parameter search}
\label{sec:effort}
Levels 1 to~4 differ in the stages that are mixed: none at level~1, stages 0 and~2 at level~2, stages 0 and 2--4 at level~3, and all five at level~4. The other stages use their direct models. Levels 5 to~9 use the model of level~4 and search for better parameters by encoding the whole file 2, 4, 6, 8, and 12 more times. The search is greedy. It first tries the two switches that are off by default, the previous-packet class context and the match model. It then tries $n_{\max}\in\{40,200,61,255,90,25,150\}$ and $\eta\in\{10,5,3\}$, in this order. Each trial changes one parameter of the best setting so far and is kept only if the archive becomes smaller. The chosen setting takes three bytes, and levels 4 to~9 decode at the same speed.

\section{Opus}
\label{sec:opus}
\subsection{Adaptive models with the codec as prior}
\label{sec:opusmodels}
Balrogg parses Opus packets with a reduced libopus decoder that keeps every branch that affects entropy coding and omits synthesis. Each call to the range decoder reports the value, the kind of symbol, its place in the syntax (the \emph{site}), and its distribution in the codec. Balrogg recodes the value with an adaptive model initialized to this distribution. Let $p_i$ be the probability of symbol $i$ in the codec, and let $S$ be a scale. A new context starts with the prior frequencies
\begin{equation}\label{eq:prior}
 f_i=\max\{1,\min\{60000,\lfloor Sp_i+1/2\rfloor\}\}
\end{equation}
and codes symbol $i$ with probability
\begin{equation}\label{eq:adapt}
 q(i)=\frac{f_i+n_i}{F+\sum_jn_j},
\end{equation}
where $F=\sum_if_i$ and the counts $n_i$ start at zero. Coding symbol $i$ adds an increment $\delta$ to $n_i$. When $\sum_jn_j$ exceeds a threshold $H$, every $n_j$ is rescaled.

Equation~\eqref{eq:adapt} is the Krichevsky--Trofimov estimator \cite{kt81} with the codec's table in place of uniform pseudo-counts.

Table~\ref{tab:params} lists the parameters and the context of each kind of symbol. Codec tables are identified by a hash of their contents, so identical tables share statistics. SILK frames contain only tabulated symbols, so the first row of Table~\ref{tab:params} covers the whole SILK layer. A binary decision that has probability $2^{-\kappa}$ in the codec uses the bit model \eqref{eq:bitupd}, initialized to that probability, with the site, $\kappa$, and the two previous outcomes at the site as context.

\begin{table}[t]
\centering
\caption{Opus multisymbol models: prior, scale $S$, increment $\delta$, rescaling threshold $H$, and context.}
\label{tab:params}
\setlength{\tabcolsep}{2.5pt}
\footnotesize
\begin{tabular}{@{}lrrrp{3.9cm}@{}}\toprule
Kind (prior)&$S$&$\delta$&$H$&Context\\\midrule
Tabulated (table)&384&28&$2^{14}$&table, site, previous value at site (clipped to 7)\\
Coarse energy (Laplace)&512&48&$2^{14}$&band, channel, intra flag, frame size; residual of previous frame, previous band, other channel (5 classes each)\\
Fine energy (uniform)&1024&16&$2^{14}$&band, channel, width; previous value; other channel (5 classes)\\
Band angle (step, triang.)&192&32&$2^{14}$&band, pdf, band-size class, resolution; previous-frame angle (8 classes)\\
Small integer (uniform)&1024&24&$2^{13}$&site, alphabet size\\
Pulse split, Eq.~\eqref{eq:pulse}&1024&16&$2^{15}$&band, depth, $(N_1,N_2,K)$\\\bottomrule
\end{tabular}
\end{table}

Coarse-energy residuals are modeled on $[-24,24]$; other values are coded as an escape symbol, a sign, and a 16-bit magnitude. Neighboring residuals enter the context in five classes: $\le-2$, $-1$, $0$, $1$, and $\ge2$. Opus writes the fine-energy bits of a band raw; Balrogg codes them as one symbol with a uniform prior, which lets the model exploit the correlation of fine energies across frames and channels. Uniform integers with a range above 1024 and the other raw bits (the signs of one-coefficient bands, the post-filter parameters, and the anti-collapse flag) are coded with a uniform distribution.

The cost of starting from the codec's table can be bounded. Every observation adds the same increment, so \eqref{eq:adapt} is a P\'olya urn. Consider one context before its first rescaling. Let $\alpha_i=f_i/\delta$ and $A=\sum_i\alpha_i$. Let a sequence of $u$ symbols contain symbol $i$ exactly $u_i$ times, and let $x^{(n)}=x(x+1)\cdots(x+n-1)$. The ideal code length of the sequence under \eqref{eq:adapt} is
\begin{equation}\label{eq:urn}
 L=\log_2\frac{A^{(u)}}{\prod_i\alpha_i^{(u_i)}},
\end{equation}
which does not depend on the order of the symbols. A static code with the codec's quantized table, $\tilde p_i=f_i/F$, has the length $L_0=\sum_iu_i\log_2(A/\alpha_i)$.

\begin{proposition}\label{prop:regret}
For every sequence of $u$ symbols coded in one context before its first rescaling,
\begin{multline*}
 L-L_0\le\frac12\sum_{i:u_i>0}\log_2\Bigl(1+\frac{u_i}{\alpha_i}\Bigr)-\frac12\log_2\Bigl(1+\frac uA\Bigr)\\
 +\log_2e\Bigl(\sum_{i:u_i>0}\frac1{12\alpha_i}+\frac1{12(A+u)}\Bigr).
\end{multline*}
\end{proposition}
\begin{IEEEproof}
By \eqref{eq:urn}, $(L-L_0)\ln2=h(A,u)-\sum_ih(\alpha_i,u_i)$ with $h(x,n)=\ln x^{(n)}-n\ln x$. Stirling's formula $\ln\Gamma(x)=(x-\frac12)\ln x-x+\frac12\ln2\pi+\mu(x)$, $0<\mu(x)<1/(12x)$, gives $h(x,n)=x\psi(n/x)-\frac12\ln(1+n/x)+\mu(x+n)-\mu(x)$ with the convex function $\psi(r)=(1+r)\ln(1+r)-r$. By Jensen's inequality, $\sum_i\alpha_i\psi(u_i/\alpha_i)\ge A\psi(u/A)$. Terms with $u_i=0$ vanish, and the bounds on $\mu$ give the claim.
\end{IEEEproof}

Let $r$ be the number of symbol values that occur in the sequence. Since $\alpha_i\ge1/\delta$, the right-hand side is at most $\frac r2\log_2(1+u\delta/r)+\frac{r\delta+1}{12}\log_2e$. Learning a context thus costs at most about $\frac12\log_2(u\delta/r)$ bits per occurring value more than the quantized table, while the gain of a model that fits the data better grows linearly in $u$. The bound is nearly tight. For a pulse split with nine possible counts and equal prior frequencies, it gives 20.3~bits over the first $H/\delta=2048$ symbols of a context, or 0.01~bit per symbol, and the worst sequence costs 20.3~bits more than the static code. The proof uses only $\alpha_i>0$, so after a rescaling the bound holds again for the next period, relative to the model's distribution at that point.

\subsection{Recursive coding of PVQ indices}
\label{sec:pvq}
CELT codes the shape of a band as the index of a vector in $\mathcal V(N,K)=\{y\in\mathbb Z^N:\sum_j|y_j|=K\}$, the integer vectors with $K$ pulses, and gives each of the $V(N,K)=|\mathcal V(N,K)|$ vectors the same probability. $V(N,K)$ grows combinatorially, so a model cannot keep a count for each index. Balrogg instead decodes the index to its vector and splits the coordinates into groups of $N_1=\lfloor N/2\rfloor$ and $N_2=N-N_1$. It codes the number $k$ of pulses in the first group with a model of the form \eqref{eq:adapt} whose prior is
\begin{equation}\label{eq:pulse}
 p(k\mid N_1,N_2,K)=\frac{V(N_1,k)\,V(N_2,K-k)}{V(N,K)},\quad0\le k\le K.
\end{equation}
The two groups, with $k$ and $K-k$ pulses, are split in the same way down to a depth limit. Below the limit, each subvector is coded by its index with a uniform distribution. The table of $V(N,K)$ covers $N,K\le208$. If the alphabet size of a band differs from the table value, the index is coded directly with a uniform distribution.

\begin{proposition}\label{prop:tree}
Let the coordinates of $\mathcal V(N,K)$ be split recursively, at any positions and down to any leaves. If every split is coded with the probabilities \eqref{eq:pulse} and every leaf by its uniform index, then every $y\in\mathcal V(N,K)$ has probability $1/V(N,K)$.
\end{proposition}
The proof is by induction over the tree, from the identity $\sum_kV(N_1,k)\,V(N_2,K-k)=V(N,K)$. For example, $y=(1,0,-1,0,0,2,0,0)\in\mathcal V(8,4)$ at depth~2 has the split probabilities $\frac{32\cdot32}{2816}$, $\frac{4\cdot4}{32}$, and $\frac{8\cdot1}{32}$ and the leaf probabilities $\frac14$, $\frac14$, $\frac18$, and $1$. Their product is $1/2816=1/V(8,4)$.

The split models start from \eqref{eq:pulse}, rounded by \eqref{eq:prior}, so all savings come from adaptation. Real band shapes are not uniformly distributed, and the models learn the distribution separately for each band, depth, and $(N_1,N_2,K)$ (last row of Table~\ref{tab:params}). The level sets the depth limit: levels 1 to~7 use depths 0 to~6, and levels 8 and~9 use depth~6.

A repeat flag precedes the tree. For each band and for coarse classes of $N$ and $K$, the coder stores the last index. If $V(N,K)>16$ and the alphabet size is the same as for the stored index, an adaptive bit signals whether the current index equals the stored one. If it does, nothing else is coded. The tree does not depend on the flag, so all vectors of a band share the split statistics.

\subsection{Exact recovery of frame bytes}
\label{sec:frame}
Encoding the decoded symbols again does not always give back the original frame. The range coder may choose some of its final bits freely. The range-coded data is written from the start of the frame and the raw bits from the end. The bytes between them are arbitrary, and FFmpeg leaves bytes of earlier frames there. The encoder and the decoder of Balrogg therefore encode the symbols with a copy of the libopus range encoder, the \emph{mirror}, into a buffer of the frame's length. Let $x$ be the original frame and $\hat x$ the output of the mirror. An adaptive flag signals whether $x\ne\hat x$. If so, the encoder stores $a=\min\{j:x_j\ne\hat x_j\}$, $b=\max\{j:x_j\ne\hat x_j\}$, and the bytes $x_a,\ldots,x_b$, and the decoder overwrites $\hat x_a,\ldots,\hat x_b$ with them. The interval can cover the whole frame, so every frame that parses can be represented.

The positions are coded as $a$ and $b-a$. Each stored byte is first predicted from the same offset in the last four frames and is coded as a literal only if none of these predictions is correct. The vendor string selects one of three encoder classes (libopus, FFmpeg, other) as the context of these models. The flag starts at probability $2^{-12}$. For FFmpeg streams, the flag starts at $1/2$, and $a$ and $b-a$ are coded as differences from the previous frame. The mirror reproduced all 2\,090\,820 frames of the MLS files of Section~\ref{sec:results} and all but 104 of the 7.6 million libopus frames of Section~\ref{sec:grid}. Table-of-contents bytes, frame counts, frame lengths, and packet padding have their own models.

\begin{table*}[t]
\centering
\caption{Compressed size (\%) of complete sets at levels 1, 4, and~9, and of xz and zstd applied to each file. Times are wall-clock seconds for the whole set at level~4 with 32 processes. Memory is the largest per-process peak in MiB at any level.}
\label{tab:results}
\begin{tabular}{llrr rrr rr rrr}\toprule
&&&&\multicolumn{3}{c}{Balrogg (\%)}&\multicolumn{2}{c}{Baselines (\%)}&\multicolumn{3}{c}{Level 4}\\
\cmidrule(lr){5-7}\cmidrule(lr){8-9}\cmidrule(lr){10-12}
Set&Codec&Files&MB&L1&L4&L9&\texttt{xz -9e}&\texttt{zstd -19}&Enc.\ (s)&Dec.\ (s)&Mem.\\\midrule
OpenMIC-2018&Vorbis&20\,000&2595.8&86.2&83.8&83.5&97.5&97.5&68.4&71.5&6.6\\
VoxPopuli&Vorbis&41&5178.9&89.8&89.4&89.3&98.6&98.3&120.6&129.7&11.7\\
\textit{Minecraft} (V)&Vorbis&741&29.5&83.9&80.8&80.4&96.9&96.9&2.4&2.5&8.6\\
Piano piece (P)&Vorbis&1&2.7&83.3&79.0&78.4&97.4&97.6&1.1&1.2&5.5\\
MLS&Opus&2729&182.1&94.9&94.6&94.3&99.5&98.6&3.4&3.5&4.7\\\bottomrule
\end{tabular}
\end{table*}

\begin{table*}[t]
\centering
\caption{Single-process comparison on identical files (OpenMIC-2018: first 1000 files; VoxPopuli: first two recordings): compressed size (\%) and throughput (MB/s), encoding/decoding.}
\label{tab:single}
\setlength{\tabcolsep}{3.5pt}
\footnotesize
\begin{tabular}{l rrrrrr cccccc}\toprule
&\multicolumn{6}{c}{Compressed size (\%)}&\multicolumn{6}{c}{Throughput, enc./dec.\ (MB/s)}\\
\cmidrule(lr){2-7}\cmidrule(lr){8-13}
Set&L1&L4&\texttt{xz -9e}&\texttt{zstd -19}&ZPAQ&bzip3&L1&L4&\texttt{xz -9e}&\texttt{zstd -19}&ZPAQ&bzip3\\\midrule
OpenMIC-2018&86.7&84.5&97.8&97.7&97.4&97.4&4.9/5.0&2.3/2.2&1.4/15.5&6.5/22.9&0.22/0.22&5.5/5.2\\
VoxPopuli&91.2&90.8&99.9&99.7&98.6&98.6&7.1/7.1&3.2/2.9&2.4/331&11.7/3320&0.33/0.33&5.9/4.9\\
Set V&83.9&80.8&96.9&96.9&97.8&96.6&3.3/3.4&1.9/1.8&1.0/6.4&4.1/7.5&0.16/0.16&3.6/3.5\\
Set P&83.3&79.0&97.4&97.6&95.9&96.2&6.0/6.3&2.5/2.4&4.6/35.8&10.7/189&0.23/0.24&6.6/6.0\\
MLS&94.9&94.6&99.5&98.6&99.6&98.8&3.7/3.9&2.7/3.0&0.9/9.8&5.5/12.4&0.19/0.19&4.5/4.4\\\bottomrule
\end{tabular}
\end{table*}

\section{Measurements}
\label{sec:results}
\subsection{Setup}
All archives were verified to decode to their inputs. Balrogg was compiled with GCC~15.2 at \texttt{-O2}. The baselines are xz~5.8.3 (\texttt{-9e}) \cite{lzma}, zstd~1.5.7 (\texttt{-19}) \cite{zstd}, ZPAQ~7.15 (\texttt{-m5}, its highest level) \cite{zpaq}, and bzip3~1.5.4 (one block per file) \cite{bzip3}, each applied to every file separately. The test machine has a 16-core, 32-thread AMD Ryzen~9 5950X processor and 64~GB of memory and runs Linux~7.0. Files are read from the page cache and written to a memory-backed file system, and each file is coded by one single-threaded process. Memory is the peak anonymous resident memory of a process.

We use five data sets. OpenMIC-2018 \cite{openmic} has 20\,000 ten-second stereo music clips in Vorbis at about 112~kb/s. The English 2020 unlabeled part of VoxPopuli \cite{voxpopuli} has 41 mono Vorbis recordings of up to 220~MB at about 48~kb/s. The Polish and Portuguese test and development splits of Multilingual LibriSpeech (MLS) \cite{mls} have 2729 Opus utterances, 15~s long on average, at about 35~kb/s. Set~V contains the 741 Vorbis sound and music files of \textit{Minecraft}~1.2.5 \cite{minecraft}, of which 449 are distinct, and set~P is one piano track from the same game. We developed the models on sets V and~P and processed all other data only after the parameters were fixed.

\subsection{Compression, speed, and memory}
At level~4 (Table~\ref{tab:results}), Balrogg reduces Vorbis music by 16.2--21.0\%, Vorbis speech by 10.6\%, and the Opus speech of MLS by 5.4\%. On the same files, xz and zstd remove 0.5--3.1\%. No file became larger. Level~9 encodes each Vorbis file 13 times and improves on level~4 by 0.1--0.6 percentage points. With 32 processes at level~4, the machine encodes OpenMIC-2018 at 38~MB/s and VoxPopuli at 43~MB/s.

\begin{table}[t]
\centering
\caption{Per-process memory in MiB, encoding/decoding, on the files of Table~\ref{tab:single}: largest peak over the files of a set.}
\label{tab:mem}
\setlength{\tabcolsep}{2.5pt}
\footnotesize
\begin{tabular}{@{}lcccccc@{}}\toprule
Set&L1&L4&\texttt{xz -9e}&\texttt{zstd -19}&ZPAQ&bzip3\\\midrule
OpenMIC&2.0/1.9&3.5/3.4&66/0.2&5.6/1.1&85/84&5.9/5.9\\
VoxPopuli&6.5/3.6&8.0/5.1&673/64&713/11&976/849&1278/1277\\
Set V&2.9/2.8&4.4/4.3&86/2.4&56/4.9&127/126&18/18\\
Set P&2.4/2.4&4.0/3.9&87/2.7&56/4.7&128/126&19/18\\
MLS&4.3/4.3&4.7/4.7&65/0.2&2.6/0.5&84/84&5.6/5.5\\\bottomrule
\end{tabular}
\end{table}

\begin{table}[t]
\centering
\caption{Per-file compressed size at level~4 (\%): median, interquartile range, 5th to 95th percentile, and range.}
\label{tab:dist}
\setlength{\tabcolsep}{4pt}
\begin{tabular}{@{}lrccc@{}}\toprule
Set&Median&IQR&5--95\%&Range\\\midrule
OpenMIC-2018&84.8&81.6--86.8&74.7--88.5&35.5--92.5\\
VoxPopuli&89.9&88.7--90.7&86.9--90.9&86.0--91.0\\
\textit{Minecraft} (V)&78.9&74.0--83.3&65.8--88.1&52.4--92.0\\
MLS&94.6&94.2--94.8&93.8--95.1&92.5--95.3\\\bottomrule
\end{tabular}
\end{table}

\begin{table}[t]
\centering
\caption{2000 files of OpenMIC-2018 (254.5~MB) at level~4: wall-clock time, speedup, and largest sum of the private memory of all running processes. Means of three runs (coefficient of variation at most 0.25\%).}
\label{tab:scale}
\begin{tabular}{@{}rrrrrr@{}}\toprule
Jobs&Enc.\ (s)&Dec.\ (s)&\multicolumn{2}{c}{Speedup}&Mem.\ (MiB)\\
\cmidrule(lr){4-5}
&&&Enc.&Dec.&\\\midrule
1&109.7&110.0&1.0&1.0&3.5\\
2&55.2&55.8&2.0&2.0&6.8\\
4&27.6&27.9&4.0&3.9&13.4\\
8&13.9&14.0&7.9&7.9&26.5\\
16&7.7&7.4&14.3&15.0&50.9\\
32&6.3&5.9&17.4&18.8&91.8\\\bottomrule
\end{tabular}
\end{table}

\begin{table*}[t]
\centering
\caption{EBU SQAM, 70 tracks per configuration: compressed size (\%) at levels 1, 4, and~9, per-file median at level~4, and \texttt{xz -9e} per file; for Opus, reduction (\%) of the entropy-coded payload at level~4. LV: libvorbis~1.3.7 (quality $q$); FF: FFmpeg~7.1 (quality $q$ for Vorbis, kb/s for Opus); LO: libopus~1.5.2 and LO$'$: libopus~1.3.1 (kb/s; stereo, 20~ms frames, and automatic mode unless stated).}
\label{tab:grid}
\setlength{\tabcolsep}{3pt}
\footnotesize
\begin{tabular}[t]{@{}lrrrrr@{}}\toprule
Configuration&L1&L4&L9&Med.&\texttt{xz}\\\midrule
LV $q{=}0$&82.1&77.5&76.3&77.8&94.4\\
LV $q{=}2$&85.5&82.2&81.4&84.1&96.1\\
LV $q{=}4$&84.7&82.4&81.8&83.6&96.0\\
LV $q{=}6$&87.8&86.2&85.9&87.4&98.0\\
LV $q{=}8$&89.1&87.7&87.4&88.8&98.4\\
LV $q{=}10$&91.5&90.5&90.3&91.8&99.1\\
LV $q{=}2$, mono&85.9&84.0&83.4&86.0&97.0\\
LV $q{=}6$, mono&88.5&86.7&86.3&88.7&97.9\\
LV $q{=}4$, 5.1&80.1&76.9&73.7&77.6&93.7\\
LV CBR 128&73.7&71.4&70.9&71.7&82.2\\
FF $q{=}2$&69.0&66.5&65.0&65.8&88.5\\
FF $q{=}5$&77.0&74.8&73.7&74.5&93.5\\
FF $q{=}8$&79.3&77.3&76.3&77.4&94.7\\
\bottomrule
\end{tabular}\hfill
\begin{tabular}[t]{@{}lrrrrrr@{}}\toprule
Configuration&L1&L4&L9&Med.&\texttt{xz}&Payl.\\\midrule
LO 16&86.1&83.5&83.3&85.9&94.2&10.6\\
LO 24&87.0&85.4&85.4&87.8&96.8&11.4\\
LO 32&89.5&88.1&88.0&90.0&97.8&9.4\\
LO 48&92.1&91.0&90.9&92.2&98.7&7.2\\
LO 64&93.4&92.5&92.2&93.5&99.0&6.0\\
LO 96&94.5&93.8&93.5&94.5&99.1&5.0\\
LO 128&95.1&94.4&94.1&95.2&99.1&4.3\\
LO 192&96.1&95.4&95.1&96.2&99.3&3.6\\
LO 256&96.8&96.1&96.0&97.1&99.2&2.8\\
LO SILK 12&78.2&78.2&78.2&78.0&90.1&13.5\\
LO SILK 16&82.6&82.6&82.6&82.4&92.4&10.6\\
LO SILK 24&83.3&83.3&83.3&83.2&93.3&12.1\\
LO SILK 32&86.3&86.3&86.3&86.2&95.2&10.0\\
LO hybrid 24&76.7&76.6&76.6&76.4&89.5&19.9\\
LO hybrid 32&80.6&80.5&80.5&80.0&91.9&16.7\\
LO hybrid 48&85.9&85.6&85.4&85.3&93.2&12.4\\
LO CELT 16&86.1&83.5&83.3&85.9&94.2&10.6\\
LO CELT 32&89.5&88.1&88.0&90.0&97.8&9.4\\
\bottomrule
\end{tabular}\hfill
\begin{tabular}[t]{@{}lrrrrrr@{}}\toprule
Configuration&L1&L4&L9&Med.&\texttt{xz}&Payl.\\\midrule
LO CELT 64&93.4&92.5&92.2&93.5&99.0&6.0\\
LO CELT 128&95.1&94.4&94.1&95.2&99.1&4.3\\
LO 24, mono&89.0&87.3&87.1&89.5&96.5&9.3\\
LO 64, mono&94.5&93.6&93.3&94.7&99.2&4.7\\
LO 64, 2.5 ms&77.0&76.8&76.8&78.3&91.9&15.4\\
LO 64, 5 ms&84.2&83.5&83.5&86.3&96.2&12.5\\
LO 64, 10 ms&90.8&89.7&89.6&91.2&98.1&7.9\\
LO 64, 40 ms&92.6&91.5&91.3&93.0&98.7&7.0\\
LO 64, 60 ms&92.6&91.5&91.3&93.0&98.8&7.0\\
LO 64, CBR&80.7&79.7&79.5&76.9&85.8&6.6\\
LO 24, speech&78.7&78.6&78.6&78.0&90.7&17.6\\
LO$'$ 32&89.5&88.1&88.0&89.9&97.8&9.4\\
LO$'$ 64&93.4&92.5&92.2&93.5&99.0&6.0\\
LO$'$ 128&95.1&94.4&94.1&95.2&99.1&4.3\\
FF 64&81.0&79.9&79.6&78.5&85.8&18.1\\
FF 128&82.4&81.4&81.1&79.2&86.1&12.7\\
FF 256&83.6&82.8&82.7&80.3&86.1&7.6\\
\bottomrule
\end{tabular}
\end{table*}

On identical files with one process (Tables~\ref{tab:single} and~\ref{tab:mem}), ZPAQ at its highest level removes 0.4--4.1\%, and bzip3 removes 1.2--3.8\%. Balrogg at level~4 removes 5.4--21.0\%, is about ten times faster than ZPAQ, and uses 3.5--8.0~MiB of memory to encode, even for 220~MB recordings.

Table~\ref{tab:dist} gives per-file distributions. Files are coded independently, so the speedup is nearly linear up to the 16 physical cores (Table~\ref{tab:scale}).

\subsection{Encoders, bit rates, and coding modes}
\label{sec:grid}
We encoded the 70 tracks of the EBU SQAM set \cite{sqam} in 48 configurations of libvorbis~1.3.7, libopus~1.5.2 and~1.3.1, and the Vorbis and Opus encoders of FFmpeg~7.1 (Table~\ref{tab:grid}). Balrogg accepted all 3360 files at levels 1, 4, and~9. At level~4, the compressed size is 66.5--90.5\% for Vorbis and 76.6--96.1\% for Opus. The reduction is largest for FFmpeg and at low Vorbis qualities. For Opus, the table also gives the reduction of the entropy-coded payload, which excludes framing, headers, and padding. With libopus, the payload shrinks by 10--20\% in SILK and hybrid modes and by 2.8--15\% in CELT mode.

\subsection{Baselines that use redundancy between files}
Table~\ref{tab:solid} adds baselines that exploit redundancy between files, such as the setup headers that libvorbis repeats in every file: a solid \texttt{tar} archive of the whole set, and zstd with a 1-MiB dictionary trained on the set, whose size is counted. Coding each file separately, Balrogg is still 11.8 and 11.7 percentage points smaller than the best of these baselines on the two Vorbis sets, and 3.5 points smaller on MLS. Balrogg can also code a concatenation of files as one chained stream, which stores repeated setup headers once and shares the models between links (last row of the table).

\begin{table}[t]
\centering
\caption{Baselines that use redundancy between files: compressed size (\%) of whole sets. Set~V is restricted to its 449 distinct files.}
\label{tab:solid}
\setlength{\tabcolsep}{4pt}
\begin{tabular}{@{}lrrr@{}}\toprule
Method&V&OpenMIC&MLS\\\midrule
\texttt{xz -9e}, per file&97.2&97.5&99.5\\
\texttt{zstd -19}, per file&97.2&97.5&98.6\\
\texttt{xz -9e}, solid archive&92.5&95.5&99.0\\
\texttt{zstd -19 --long=31}, solid archive&92.5&95.6&98.1\\
\texttt{zstd -19}, trained dictionary&92.7&95.5&98.3\\\midrule
Balrogg level 4, per file&80.7&83.8&94.6\\
Balrogg level 4, chained stream&77.4&82.8&93.6\\\bottomrule
\end{tabular}
\end{table}

Against OptiVorbis~0.3.2 (Table~\ref{tab:ov}), Balrogg saves 2.6 to 5.8 times as much, and it also compresses the output of OptiVorbis.

\begin{table}[t]
\centering
\caption{Compressed size (\%): OptiVorbis, Balrogg at level~4, and Balrogg at level~4 applied to the output of OptiVorbis.}
\label{tab:ov}
\begin{tabular}{@{}lrrr@{}}\toprule
Set&OptiVorbis&Balrogg&Both\\\midrule
OpenMIC-2018&94.5&83.8&83.5\\
VoxPopuli&97.1&89.4&89.4\\
\textit{Minecraft} (V)&92.7&80.8&78.7\\
Piano piece (P)&96.4&79.0&79.0\\\bottomrule
\end{tabular}
\end{table}

\begin{table}[t]
\centering
\caption{Internet Archive files at level~4: files examined and accepted, compressed size of the accepted files (\%) with a 95\% bootstrap interval over files, and per-file median.}
\label{tab:ia}
\setlength{\tabcolsep}{3pt}
\footnotesize
\begin{tabular}{@{}lrrrcr@{}}\toprule
Collection&Files&Acc.&Size&95\% interval&Med.\\\midrule
\multicolumn{6}{@{}l}{\emph{Opus, random sample}}\\
All files&1002&1000&94.1&94.0--94.2&94.3\\
\quad FFmpeg&837&837&94.1&93.9--94.2&94.3\\
\quad libopus&107&107&94.6&94.0--95.1&95.2\\
\quad WhatsApp&35&34&91.9&91.5--92.4&92.5\\
\quad Other encoders&23&22&91.3&88.4--93.8&88.5\\\midrule
\multicolumn{6}{@{}l}{\emph{Vorbis, random sample, uploaded files}}\\
All files&493&468&85.0&84.3--85.7&87.1\\
\quad libvorbis 2001--2007&126&119&85.2&83.7--86.5&85.9\\
\quad libvorbis 2009--2020&264&246&84.9&84.0--85.8&87.1\\
\quad FFmpeg&79&79&84.2&82.7--85.7&88.6\\
\quad aoTuV&22&22&88.0&84.3--90.8&90.1\\
\multicolumn{6}{@{}l}{\emph{Vorbis, random sample, archive-derived files}}\\
All files&1011&880&83.9&83.6--84.3&84.4\\\midrule
\multicolumn{6}{@{}l}{\emph{Opus, earlier collection (distinct files)}}\\
All files&3727&3725&95.2&95.1--95.2&95.6\\
\quad One podcast&2848&2848&95.4&95.4--95.4&95.7\\
\quad Other items&879&877&93.8&93.4--94.2&94.6\\\bottomrule
\end{tabular}
\end{table}

\subsection{Files from the Internet Archive}
We drew random samples from the Internet Archive with a fixed seed. The Opus sample is the first uploaded \texttt{.opus} file of each of 1002 audio items. The Vorbis samples come from Creative Commons items with a Vorbis file: 493 uploaded files and 1011 files that the archive derived with its own conversion service. An earlier collection of 3727 distinct Opus files is listed separately, because 2848 of its files belong to one podcast. No file became larger (Table~\ref{tab:ia}). The compressed size is 94.1\% for the Opus sample and 83.9--85.0\% for the Vorbis samples, which include files written by 17 releases of libvorbis from 2001 to 2020. All files not accepted have damaged containers (an incomplete last page, or bytes outside any page).

\section{Conclusion}
We presented Balrogg, a lossless recompressor for Vorbis and Opus. For Vorbis, it codes centered multiplicand digits with a logistic mixer that starts at the direct model and selects its weights by the neighboring digits and the previous residue pass. For Opus, it codes each symbol with an adaptive model whose prior is the codec's table, at a learning cost of about $\frac12\log_2 u$ bits per symbol value, and splits PVQ indices recursively without changing the codec's code length. Balrogg reduces Vorbis files by 10.6--21.0\% and Opus speech by 5.4\% on published data sets, by 3.9--33.5\% over 48 encoder configurations, and by 5.9\% (Opus) and 15.0--16.1\% (Vorbis) on random samples from the Internet Archive. It runs at 1.8--3.2~MB/s per core in both directions and uses less than 12~MiB of private memory.

\end{document}